# Nonvolatile memory effects in hybrid devices of few-layer graphene and ferroelectric polymer films

Yong-Joo Doh[1] and Gyu-Chul Yi[2,a]

[1]*Department of Physics, POSTECH, Pohang, Gyungbuk 790-784, Republic of Korea*

[2]*Department of Physics and Astronomy and National Creative Research Initiative Center for Semiconductor Nanorods, Seoul National University, Seoul 151-747, Republic of Korea*

ABSTRACT

We report on the fabrication and electrical characterization of few-layer graphene (FLG) devices coated with a ferroelectric polymer layer of poly(vinylidene fluoride/trifluoroethylene) [P(VDF/TrFE)]. Highly stable and reliable resistance changes were observed under floating conditions, which were dependent on the back gate voltage applied beforehand. Nonvolatile memory functionality in the hybrid FLG-P(VDF/TrFE) devices is attributed to a remanent electric field induced by the ferroelectric polarization of the P(VDF/TrFE) layer.

[a] Authors to whom correspondence should be addressed; E-mail: gcyi@snu.ac.kr

The successful isolation of high-quality graphene,[1] which is essentially a single layer of graphite, has opened up a new field of materials science and low-dimensional physics. The remarkable electronic properties of graphene, which include a high carrier mobility in excess of 20,000 cm$^2$/Vs and a micrometer-scale mean free path at room temperature,[2] have stimulated various efforts to develop novel electronic devices based on "quasi-relativistic" carrier dynamics.[3,4] Thus far, this technology has yielded a high sensitivity and low-noise chemical detector,[5] nanoribbon transistor,[6] quantum dots,[7] and logic gates.[8] In particular, recent advancements in large-scale growth of mono- and few-layer graphene (FLG) films on arbitrary substrates[9,10] has facilitated the development of graphene electronics integrated on a transparent and/or flexible substrate.

In this letter, we report the fabrication and electrical characterization of FLG devices passivated by a ferroelectric polymer layer. Poly(vinylidene fluoride/trifluoroethylene) [P(VDF/TrFE)] copolymer layers[12] are widely employed in organic ferroelectric field-effect transistors.[13-16] In the current study, P(VDF/TrFE) films were spin-coated onto the FLG, which provided for intimate contact with the FLG layer and simplified fabrication by utilizing solution processing at low temperature. Highly stable and reproducible resistance changes were observed in the polymer-combined FLG devices depending on the application history of the back gate voltage. This is attributed to ferroelectric polarization switching of the P(VDF/TrFE) layer. The observed memory effect when combined with the ferroelectric polymer layer makes these materials potentially useful in organic nonvolatile memory devices. Furthermore, the use of graphene or FLG thin films makes these materials compatible with both glass and plastic substrates.

FLG films were mechanically exfoliated from graphite flakes (NGS Naturgraphit GmbH) onto a highly doped silicon substrate covered with a 300-nm-thick $SiO_2$ layer. A silicon substrate was used as the back gate electrode. Electrical contacts were formed by using standard electron-beam lithography and subsequent electron-beam evaporation of Ti (5 nm) and Au (50 nm). A schematic diagram and optical microscopy image of a typical bare FLG device are shown in the insets of Figs. 1(a) and (b), respectively. Topographic images obtained on an atomic force microscope (Nanoscope IV, Veeco Inst.) revealed that the overall thickness of the FLG film between the source and drain electrodes ranged from 1 to 4 nm. A solution of P(VDF/TrFE) (65/35 mol% copolymer, Solvay Chemicals) in 2-butanone (6 wt%) was spin-coated onto the FLG film at 3000 rpm and annealed on a hot plate for 2 h at 145°C in ambient air to enhance the crystallinity of the P(VDF/TrFE) layer. The thickness of the P(VDF/TrFE) film, determined by a scanning electron microscopy image of the sample cross section, was approximately 440 ± 15 nm.

All current–voltage ($I$–$V$) and resistance–gate voltage ($R$–$V_g$) measurements were performed with a two-point configuration in ambient air at room temperature using a DC voltage source (NI-DAQ, National Instruments Co.), current preamplifier (Model 1211, DL Inst.), and a high voltage source (Keithley 2400). Capacitance–voltage ($C$–$V$) measurements were performed with a precision LCR meter (HP 4284A, Agilent Technologies) and a semiconductor characterization system (4200-SCS, Keithley Instruments Inc.) with an AC signal frequency and amplitude of 100 kHz and 500 mV, respectively. A schematic diagram of the experimental setup for $C$–$V$ measurements is depicted in Fig. 2.

Figure 1(a) shows a typical, linear $I$–$V$ curve obtained on the bare FLG device. The corresponding $R$–$V_g$ curve in Fig. 1(b) displays a resistance peak of $R$ = 2.8 k$\Omega$, or in terms of square resistance, $R_{sq}$ = 1.8 k$\Omega$ near the "charge neutrality" point of $V_g$ = 48

V.[17] On data obtained from more than 12 FLG devices, the square resistance peak values ranged from 0.5 to 3.1 k$\Omega$ at their respective "charge neutrality" points occurring between $V_g$ = 33 and 68 V. Contact resistance was approximately 40 $\Omega$ when measured using a three-probe configuration on the bare device, corresponding to 1.1 $\mu\Omega$cm$^2$ as the product of resistance and contact area.

To characterize the ferroelectric properties of the P(VDF/TrFE) layer, $C$–$V$ measurements were performed with capacitors containing the P(VDF/TrFE) film as the dielectric medium, as shown in Fig. 2. During voltage sweeps between –40 and 40 V, a so-called "butterfly" $C$–$V$ curve was observed, which is characteristic of a ferroelectric phase transition in the P(VDF/TrFE) layer.[15,18] The curve exhibits two capacitance peaks near $V_g$ ~±25 V as a result of ferroelectric polarization switching. The coercive electric field and the dielectric constant of the P(VDF/TrFE) film were ~0.6 MV/cm and 10.8–11.3, respectively, as estimated from the switching voltage and the capacitance. Similar results were obtained on an additional five devices and are comparable to previous reports.[15,19] Discrepancies with previous studies are likely due to differences in the specific mole fraction of the copolymer and different annealing conditions.

After spin-coating the P(VDF/TrFE) layer onto the surface of the FLG film, the $R$–$V_g$ curve showed a highly hysteretic behavior in the clockwise direction with $V_g$ sweeps in the negative and positive directions. Hysteresis in the $R$–$V_g$ curve was particularly enhanced at slower sweep rates and larger $V_g$ ranges, as shown in Figs. 3(a) and (b). Similar behavior has been observed in thin film transistors with polymeric ferroelectric gates.[14,15] In contrast, the hysteresis was nearly negligible at the high scan rate of 5 V/s. This implies that the ferroelectric switching of the P(VDF/TrFE) layer is not fast enough to track the high $V_g$ sweep rate.

Note that the hysteresis in Fig. 3(b) is almost negligible at $V_g$ sweep ranges below 30 V. The inset shows that the resistance difference ratio, $R_{ratio} = \Delta R/R_{avg}$, where $\Delta R$ is the difference between and $R_{avg}$ is the mean value of two different $R$ values at $V_g = 0$, displays a sudden jump at a sweep range of 30 V and thereafter increases with the amplitude of $V_g$. This feature indicates that the coercive voltage for ferroelectric switching in the P(VDF/TrFE) layer is close to 30 V, which is consistent with the results of $C$–$V$ measurements (Fig. 2). In addition, the sawtooth-like $R$–$V_g$ curve in Fig. 3(b) was attributed to localized nucleation of ferroelectric domains with reversed polarization,[20] and could be smoothed by repeating the $V_g$ sweep.

The hysteresis of the $R$–$V_g$ curve obtained with these hybrid FLG-P(VDF/TrFE) devices can be utilized to perform nonvolatile memory functionality, which requires relatively long retention properties. Figure 4(a) shows two $I$–$V$ curves measured at $V_g = 0$ following a 20-s application of $V_g = 70$ or $-70$ V. The linear resistance, $R$, obtained under floating conditions ($V_g = 0$), was strongly dependent on the memory of $V_g$, which had been applied previously. Practical performance of these hybrid devices can be characterized by $R_{ratio}$, as listed in Table I. The highest resistance modulation ratio was $R_{ratio} \sim 11\%$. This can be explained by the intrinsic semimetallic nature of graphene and FLG films and implies the absence of an energy gap, which indicates the absence of an insulating state.

Temporal measurements of $R$ are shown in Fig. 4(b). Here, the hybrid FLG-P(VDF/TrFE) device was biased at $V_g = 70$ or $-70$ V for 20 s and the time-dependent drain current was measured at $V_g = 0$ with a constant drain voltage of $V = 0.1$ V. Most importantly, the memory effect was retained for a relatively long period of time with a $R_{ratio}$ of 5% even after 30 min. From an exponential fit of the time-dependent $\Delta R$, the $R_{ratio}$ was estimated to approximate zero after 180 h. Such a long retention suggests a different mechanism from transient behaviors caused by charge trapping, charge injection, or ion migration.[14] Rather, the memory functionality may be attributed to ferroelectric polarization of the P(VDF/TrFE) layer, which was also responsible for the hysteretic $R$–$V_g$ curve. This suggests that the channel resistance of the underlying FLG film can be modulated by surface charges induced at the interface between FLG and the P(VDF/TrFE) films as a result of ferroelectric polarization.

Figure 4(c) demonstrates reliable operation of the nonvolatile memory effect in hybrid FLG-P(VDF/TrFE) devices subjected to a sequence of $V_g$ applications. The resistance at $V_g = 0$ alternates between $R = 3.5$ and $4.0$ k$\Omega$ as a result of $V_g$-induced polarization of the ferroelectric layer. When $V_g$ was switched from 0 to 70 or –70 V, a sharp resistance peak was observed as a result of transient displacement current due to the polarization change of the ferroelectric layer.[21] As a control, the resistance response from a monotonous $V_g$ pulse is shown in Fig. 4(d), which supports the idea that the observed $V_g$-induced memory effect was caused primarily by ferroelectric polarization switching rather than charge trapping or chemical modification of the graphene.[22]

In summary, hybrid devices composed of FLG and ferroelectric polymer films were fabricated and electrically characterized. These devices exhibited nonvolatile memory retention properties under ambient conditions with relatively long retention times and reliable operation under pulsed $V_g$. The primary mechanism for this memory effect was attributed to the ferroelectric polarization of the P(VDF/TrFE) layer. The observed memory effect in the hybrid FLG devices with the ferroelectric polymer layer makes these materials potentially useful in transparent and flexible electronics.

The authors would like to acknowledge valuable discussions with H.-J. Lee and technical assistance from J.-H. Choi and S.-W. Ryu. This work was financially supported by the National Creative Research Initiative Project (R16-2004-004-01001-0) of the Korea Science and Engineering Foundation (KOSEF) and by Acceleration Research (R17-2008-007-01001-0).

TABLE I. The physical dimensions and device performance parameters are given for hybrid FLG-P(VDF/TrFE) devices. $L$ is the separation between source and drain electrodes, and $w$ represents the overall width of the FLG film. $\Delta R$, $R_{avg}$, and $R_{ratio}$ are explained in the text.

| No. | $L$ ($\mu$m) | $w$ ($\mu$m) | $\Delta R$ ($\Omega$) | $R_{avg}$ ($\Omega$) | $R_{ratio}$ (%) |
|---|---|---|---|---|---|
| 1 | 2.8 | 1.9 | 48 | 594 | 8.4 |
| 2 | 3.0 | 0.7 | 450 | 4390 | 10.8 |
| 3 | 3.7 | 2.0 | 38 | 1070 | 3.6 |
| 4 | 2.8 | 1.1 | 132 | 1940 | 7.0 |

**FIGURE CAPTIONS**

FIG. 1. (a) A current–voltage (*I*–*V*) curve obtained on a bare FLG device at room temperature in air is shown. The back gate voltage was set to $V_g = 0$. The inset shows a schematic diagram of the FLG device. (b) The linear resistance versus the back gate voltage (*R*–$V_g$) is shown at $V = 0.1$ V for the same device. The charge-neutrality peak was observed near $V_g = 48$ V. The inset shows an optical microscope image of the device. The scale bar represents 5 $\mu$m.

FIG. 2. The capacitance–voltage (*C*–*V*) curve is shown for the P(VDF/TrFE) layer. The observed double hysteresis was due to ferroelectric switching of the P(VDF/TrFE) layer. A schematic diagram illustrates the capacitance measurement configuration.

FIG. 3. (a) The *R*–$V_g$ curves were obtained while varying the sweep rate of $V_g$ on hybrid device No. 2. The clockwise hysteresis was most clearly observed at slower sweep rates. The inset shows a schematic diagram of the hybrid device composed of FLG and the polymeric ferroelectric film and its measurement configuration. (b) The *R*–$V_g$ curve from hybrid device No. 3 is shown with a sweep ranges of $V_g = \pm 20$ (black), $\pm 40$ (blue), and $\pm 70$ (red) V. The inset shows how the resistance modulation ratio at $V_g = 0$ varied as a function of the $V_g$ sweep range. A bias voltage of $V = 0.1$ V was applied in both (a) and (b) and a sweep rate of 1 V/s was chosen for (b).

FIG. 4. (a) The *I*–*V* characteristics of hybrid device No. 4 are shown by solid and open circles under floating gate conditions ($V_g = 0$) in which back gate voltages of 70 or –70 V were applied previously for 20 s, respectively. (b) The corresponding retention profile was measured at $V_g = 0$ with $V = 0.1$ V. (c) Temporal measurements of *R* (red line) are shown while alternating the amplitude of $V_g$ pulses between 70 and –70 V with an interval and duration time of 60 s. The sequence of $V_g$ applications (black) is shown

schematically at the bottom of the figure.  (d) The same plot was regenerated using monotonous $V_g$ pulses with an amplitude of $V_g = 70$ V and an interval and duration time of 30 s, respectively. The bias voltage was constant at $V = 0.1$ V.

**Fig. 1**

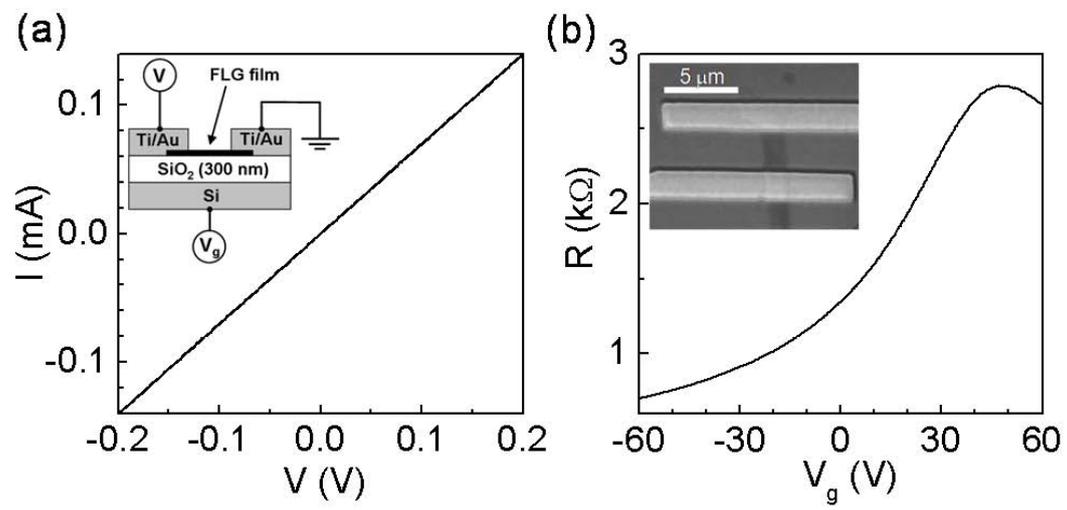

**Fig. 2**

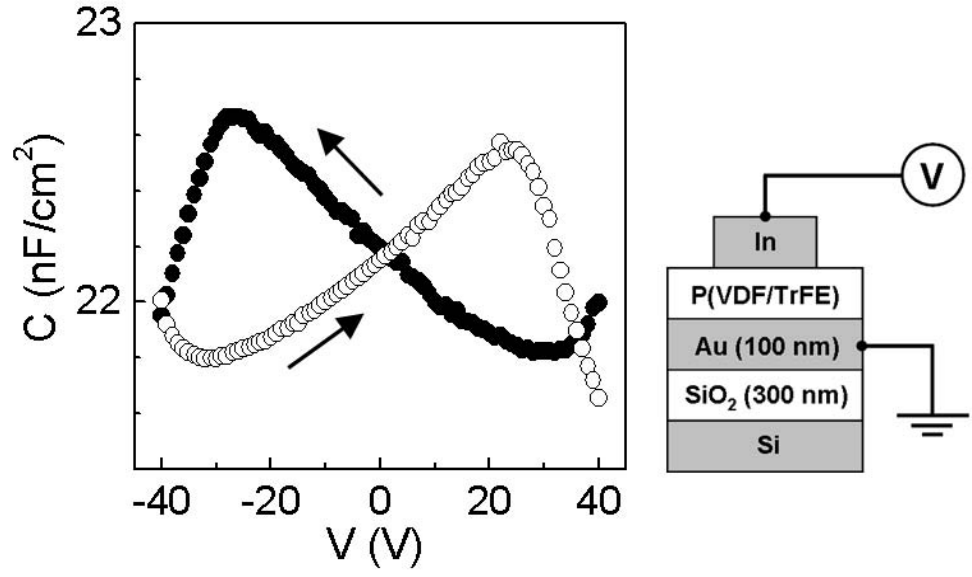

**Fig. 3**

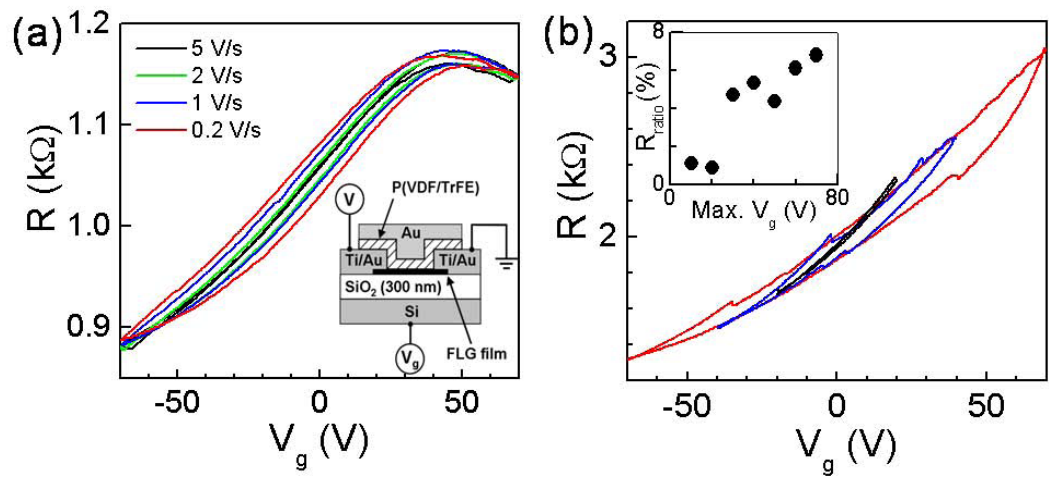

Fig. 4

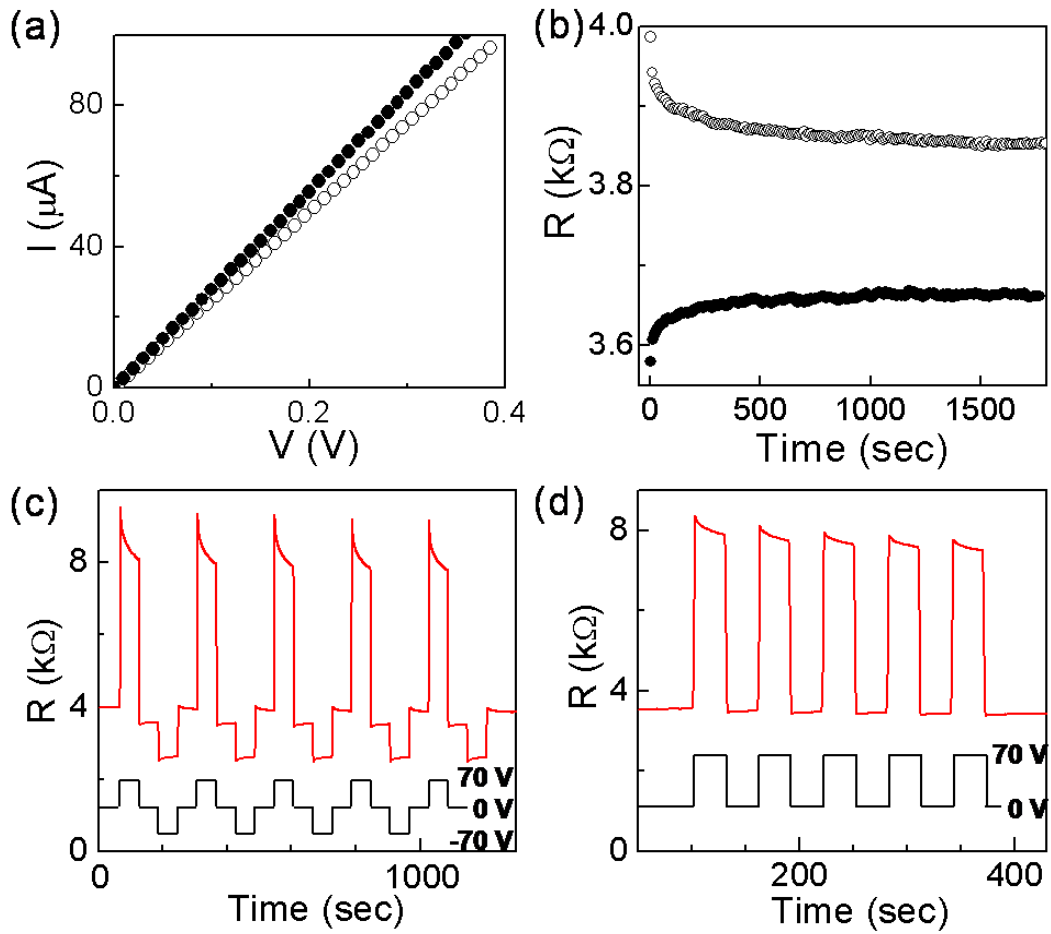